\pdfoutput=1
\documentclass[runningheads]{llncs}
\usepackage{graphicx}
\usepackage{comment}
\usepackage{amsmath,amssymb} 
\usepackage{color}
\usepackage{pifont}
\usepackage{subfigure}
\usepackage{booktabs}
\usepackage{array}
\usepackage{multirow}
\usepackage[pagebackref=true,breaklinks=true,colorlinks,bookmarks=false]{hyperref}

\usepackage[width=122mm,left=12mm,paperwidth=146mm,height=193mm,top=12mm,paperheight=217mm]{geometry}

\graphicspath{{figures/}}

\newcommand{\PreserveBackslash}[1]{\let\temp=\\#1\let\\=\temp}
\newcolumntype{C}[1]{>{\PreserveBackslash\centering}p{#1}}
\newcolumntype{R}[1]{>{\PreserveBackslash\raggedleft}p{#1}}
\newcolumntype{L}[1]{>{\PreserveBackslash\raggedright}p{#1}}

\begin{document}
\pagestyle{headings}
\mainmatter
\def\ECCVSubNumber{7323}  

\title{FC$^2$N: Fully Channel-Concatenated Network for Single Image Super-Resolution} 

\titlerunning{FC$^2$N for Single Image Super-Resolution}
%
\author{\small Xiaole Zhao\inst{1}\and
Ying Liao \inst{1}\and
Tian He\inst{1}\and
Yulun Zhang\inst{2}\and
Yadong Wu\inst{3}\and
Tao Zhang\inst{1}}
\authorrunning{Xiaole Zhao et al.}
%
\institute{School of Life Science and Technology, UESTC, Chengdu, China \\ \email{zxlation@foxmail.com; tao.zhang@uestc.edu.cn}\and
Department of ECE, Northeastern University, Boston, USA \\ \email{yulun100@gmail.com}\and
Sichuan University of Science and Engineering, Yibin, China}
\maketitle

\begin{abstract}
Most current image super-resolution (SR) methods based on convolutional neural networks (CNNs) use residual learning in network structural design, which favors to effective back propagation and hence improves SR performance by increasing model scale. However, residual networks suffer from representational redundancy by introducing identity paths that impede the full exploitation of model capacity. Besides, blindly enlarging network scale can cause more problems in model training, even with residual learning. In this paper, a novel fully channel-concatenated network (FC$^2$N) is presented to make further mining of representational capacity of deep models, in which all interlayer skips are implemented by a simple and straightforward operation, weighted channel concatenation (WCC), followed by a $1\times1$ conv layer. Based on the WCC, the model can achieve the \textit{joint} attention mechanism of linear and nonlinear features in the network, and presents better performance than other state-of-the-art SR models with fewer model parameters. To our best knowledge, FC$^2$N is the first SR model that does not adopt residual learning and reaches network depth over \textbf{400} layers. Moreover, it shows excellent performance in both largescale and lightweight implementations, which illustrates the full exploitation of the representational capacity of the model.
\end{abstract}

\section{Introduction}
\vspace{-2mm}
Single image super-resolution (SISR) is a classic problem in low-level computer vision that aims at reconstructing a high-resolution (HR) image from one single low-resolution (LR) image. Although a lot of solutions have been proposed for image SR, it is still an active yet challenging research topic in computer vision community due to its ill-poseness nature and high practical values \cite{Baker2002Limits,Zhang2018Learning,Zhao2019Channel}.

In recent years, deep learning techniques \cite{Lecun2015Deep}, especially convolutional neural networks (CNNs) \cite{Lecun1989Backpropagation,LeCun1989Handwritten} and residual learning \cite{He2016Deep}, have significantly promoted the advance of image SR. A representative work that successfully adopts CNNs to SR problem is SRCNN \cite{Dong2016Image}, which is a three-layer CNN network that can learn an end-to-end mapping between LR and HR images and achieve satisfactory SR performance at that time. Subsequently, many studies were conducted to design and build more accurate and efficient SR networks, such as \cite{Kim2016Deeply}, \cite{Tai2017Image}, \cite{Kim2016Accurate}, \cite{Zhao2019Channel}, \cite{Lim2017Enhanced}, \cite{Zhang2018Residual}, \cite{Zhang2018Image} etc. One of the major trends in these models is models get larger and deeper for further performance gain. As model scale increases, the training difficulty caused by information weakening and over/underfitting becomes more serious, and more tricks are needed to ensure an effective training \cite{Li2018Multi}. Residual learning \cite{He2016Deep} is probably one of the most commonly-used techniques to ease this training difficulty, which is a simple element-wise addition of features at different layers. It can help extend deep SR models to previously unreachable depths and capabilities by introducing identity paths that carries gradients throughout the extent of very deep models \cite{Veit2016Residual}. These identity paths, however, result in a large amount of representational redundancy in deep residual networks \cite{Huang2016Deep}, hindering the full mining of model capabilities.

\begin{figure}[t]
  \centering
  \includegraphics[width=0.99\textwidth]{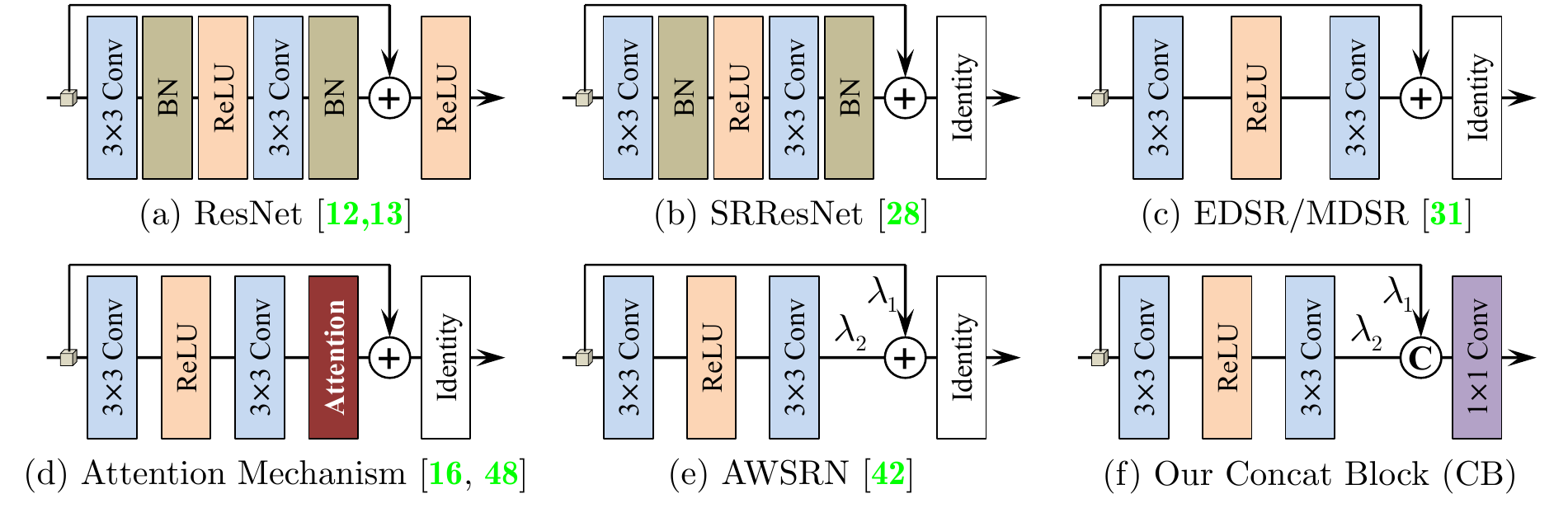}
  \vspace{-3mm}
  \caption{Typical building blocks. ``Identity'' and ``Attention'' represent identity mapping and attention mechanism respectively, and ``\textbf{C}'' denotes channel concatenation.}
  \label{fig:building_blocks}
\end{figure}

To make full use of features and further mine the representational capacity of deep networks, many SR models tend to combine residual learning with channel concatenation, which is believed to help new feature exploration and learning good representations \cite{Huang2017Densely,Chen2017Dual,Hu2018Channel,Zhao2019Channel}, e.g., MemNet \cite{Tai2017Memnet}, AWSRN \cite{Wang2019Lightweight}, RDN \cite{Zhang2018Residual}, MSRN \cite{Li2018Multi}, CARN \cite{Ahn2018Fast}, CSSFN \cite{Zhao2019Single} and DBDN \cite{Wang2018Deep} etc. On the one hand, most of these models utilize residual learning \cite{He2016Deep,He2016Identity} for stable and effective model training, but as mentioned above, it is not conducive to full exploration of model capacities. On the other hand, channel concatenation in these models is adopted to directly connect different layers, which ignores the contribution of adaptive connection strength to model capacities, i.e., weighted channel concatenation (WCC). Although the weighted interlayer connections are considered to impede effective back propagation in case of residual learning \cite{He2016Identity}, we believe that they are more in line with the manner neurons behave in the human brain, therefore more physiologically sound.

Considering these problems, we present a novel architecture for single image SR tasks in this work, in which all skip connections are simply implemented by channel concatenation without any residual connection. Each branch of these channel concatenations is attached by a weighting factor to further explore the representational capacity of the model. As shown in Fig.\ref{fig:building_blocks} and Fig.\ref{fig:FC2N}, we construct the building modules of our model by this WCC, including concat block (CB) for effective local feature utilization and concat group (CG) for the ease of model training. Moreover, we improve the global feature fusion (GFF) \cite{Zhang2018Residual} by weighting the branches of GFF's skip connections, i.e., WGFF. These structural ingredients built with our WCC can combine linear and nonlinear features from fine to coarse grained, allowing us to build deeper and more expressive models without residual learning. Since all interlayer skips are conducted by channel concatenation, the proposed model is terms as fully channel-concatenated network (FC$^2$N).

In summary, the main contributions of this work are four-fold: (1) We present a novel FC$^2$N for efficient and accurate image SR, in which channel concatenation is used to conduct all interlayer skips. It can reach much deeper without residual learning, and gives better performance boosting with fewer parameters. (2) We propose a new interlayer connection, i.e., WCC, for efficient feature fusion. It is simple and can be easily embedded into other models for performance gain. We demonstrate that residual block \cite{Lim2017Enhanced} and weighted residual block \cite{Wang2019Lightweight} are special cases of our CB. (3) We improve the GFF strategy \cite{Zhang2018Residual} by our WCC, exploring a more efficient WGFF that makes the utilization of hierarchical features more flexible and reasonable, thus promoting the model performance. (4) We construct a CG module to build a very deep and trainable network with a concat in concat (CIC) structure, which is similar to the residual in residual (RIR) structure in RCAN \cite{Zhang2018Image} but allows features are integrated in a more effective way.

\vspace{-2mm}
\section{Related Work}
\vspace{-2mm}
\subsection{Image Super-Resolution with Deep Learning}
\vspace{-2mm}
The pioneering work that utilizes deep learning techniques to solve single image super-resolution tasks in the modern sense is SRCNN \cite{Dong2016Image}, which is a three-layer network that maps LR images to HR images in an end-to-end manner. Through introducing global residual learning (GRL), Kim \textit{et al}. \cite{Kim2016Accurate} increased the network depth up to 20 layers and achieved significant performance improvement. Tai \textit{et al}. \cite{Tai2017Memnet} presented a very deep memory network (MemNet) to solve the problem of long-term dependency. Instead, some other works, e.g, DRCN \cite{Kim2016Deeply} and DRRN \cite{Tai2017Image}, focused on weight sharing to reduce the scale of model parameters. Although these methods achieve superior performance, they use the bicubic-interpolated version of original LR images as input, which inevitably loses some details and increases computational burden greatly \cite{Shi2016Real,Zhang2018Image}.

This problem can be alleviated by placing nonlinear inference in LR image space and upscaling image resolution at the end of the network, such as transpose convolution \cite{Dong2016Accelerating} and ESPCNN \cite{Shi2016Real}. Benefitting from this, some SR models can improve performance by significantly increasing their network scale, e.g., EDSR \cite{Lim2017Enhanced}, RDN \cite{Zhang2018Residual}, D-DBPN \cite{Haris2018Deep} and RCAN \cite{Zhang2018Image} etc. However, the performance gain of these methods depends largely on the increase of model scale, e.g., EDSR \cite{Lim2017Enhanced} has about 43M model parameters and 70-layer network depth, and RCAN \cite{Zhang2018Image} also has more than 16M parameters and 400-layer network depth.

To generate more realistic results, especially for large scaling factors, Ledig \textit{et al}. \cite{Ledig2017Photo} proposed to introduce generative adversarial network (GAN) \cite{Goodfellow2014Generative} into image SR framework (SRGAN). They developed a new network structure based on ResNet \cite{He2016Deep,He2016Identity} (SRResNet) and treated it as the generator of a GAN \cite{Goodfellow2014Generative} with a perceptual loss \cite{Johnson2016Perceptual}. The idea was also introduced into EnhanceNet \cite{Sajjadi2017EnhancedNet} that combined automated texture synthesis and perceptual losses. However, although these GAN-based models can ease over-smoothing artifacts and present visually pleasuring results, their results may not be faithfully recovered with wrong image content and unpleasing artifacts \cite{Lai2017Deep,Zhang2018Image}.

\vspace{-3mm}
\subsection{Interlayer Bypass Connections}
\vspace{-1mm}
A simple and direct way to improve the performance of deep models is to increase model scale, e.g., network parameters, depth and width. However, more problems will arise as model scale increases, and more training tricks are needed to ensure effective training \cite{Li2018Multi}. To alleviate the training difficulty caused by the increased model scale, interlayer connections are widely used in network design. Residual connection \cite{He2016Deep,He2016Identity} and channel concatenation \cite{Huang2017Densely} are two typical interlayer skip connections. Although residual connection is a commonly used option in image SR, e.g., \cite{Kim2016Accurate}, \cite{Tai2017Image}, \cite{Tai2017Memnet}, \cite{Lim2017Enhanced} and \cite{Zhang2018Image}, there is a large amount of representational redundancy in residual networks \cite{Huang2016Deep,Tong2017Image}, which hints that residual learning may hinder the full mining of model capacities. In fact, when model scale is relatively fixed, the performance of residual models still has potential to be improved \cite{Ahn2018Fast,Wang2019Lightweight}.

Channel concatenation is another way to implement skip connections in the context of image SR, such as MemNet \cite{Tai2017Memnet}, SRDenseNet \cite{Tong2017Image}, RDN \cite{Zhang2018Residual}, MSRN \cite{Li2018Multi}, as well as AWSRN \cite{Wang2019Lightweight} etc. However, these models usually combine channel concatenation with residual connections, expecting to make full use of features and mitigate the training difficulty. Instead, in the proposed FC$^2$N, all interlayer connections are implemented by our WCC. In this way, the model can adaptively build a reasonable amount of interlayer connections with appropriate strength and thereby make full exploitation of the representational capacity.

\vspace{-3mm}
\section{Fully Channel-Concatenated Network}
\vspace{-2mm}
\subsection{Weighted Channel Concatenation}
\vspace{-2mm}
Most current deep models are modularized architectures that consist of many stacked building blocks, e.g., ResNet \cite{He2016Deep}, MemNet \cite{Tai2017Memnet}, DRRN \cite{Tai2017Image}, SRResNet \cite{Ledig2017Photo}, EDSR \cite{Lim2017Enhanced}, AWSRN \cite{Wang2019Lightweight}, RCAN \cite{Zhang2018Image} etc. The structure of some typical building blocks is outlined in Fig.\ref{fig:building_blocks}. In the context of image SR, Conv-ReLU-Conv based residual block \cite{Lim2017Enhanced} and its variants are broadly adopted as the building modules of deep SR models, such as Fig.\ref{fig:building_blocks}(d) and Fig.\ref{fig:building_blocks}(e). Most these building blocks, however, are combined with the strategy of residual learning for efficient model training.

The building block of our FC$^2$N model is also based on the Conv-ReLU-Conv structure, but it avoids using residual learning. Instead, we adopt WCC followed by a 1$\times$1 conv layer to integrate the input and output of the Conv-ReLU-Conv structure, as shown in Fig.\ref{fig:building_blocks}(f). Temporarily, let $\mathbf{x}_{t} \in \mathbb{R}^{H \times W \times C}$ be the input of a CB and $\mathcal{H}(\cdot)$ the function corresponding to the nonlinear mapping branch, the WCC can be formulated as:
\begin{equation}\label{eqn:WCC}
  \mathbf{x}_{t+1} = \mathcal{L}([\lambda_{1}\mathbf{x}_{t}, \lambda_{2}\mathcal{H}(\mathbf{x}_{t})]),
\end{equation}
where $\mathbf{x}_{t+1}$ is the output of the CB, and $[\ldots]$ represents the operation of channel concatenation. $\mathcal{L}(\cdot)$ denotes the 1$\times$1 conv, and $\lambda_{1}$ and $\lambda_{2}$ are the corresponding weighting factors, as shown in Fig.\ref{fig:building_blocks}(f).

Let's rewrite $\mathbf{x}_{t} = [x_{t}^{1}, \ldots, x_{t}^{i}, \ldots, x_{t}^{C}]$ and $\mathcal{H}(\mathbf{x}_{t}) = [x_{h}^{1}, \ldots, x_{h}^{j}, \ldots, x_{h}^{C}]$, both of which have $C$ feature maps with spatial size of $H \times W$. Denote the kernel of 1$\times$1 conv as $\mathbf{K} \in \mathbb{R}^{2C \times C}$ with $2C$ input channels and $C$ output channels\footnote[1]{In implementation, the shape of $\mathbf{K}$ is $[k, k, C_\text{in}, C_\text{out}]$, where $k$ is kernel size, $C_\text{in}$ and $C_\text{out}$ are the number of input and output channels. Since the spatial size of $\mathbf{K}$ is 1$\times$1 here, we remove the singular dimensions for simplification.}, and omit the biases, then $\mathbf{x}_{t+1}$ is obtained by:
\begin{equation}\label{eqn:channel_concat}
  {x}_{t+1}^{u} = \sum_{i = 1}^{C}\lambda_{1}x_{t}^{i}\mathbf{K}(i, u) + \sum_{j=1}^{C}\lambda_{2}x_{h}^{j}\mathbf{K}(C+j,u),
\end{equation}
where $x_{t+1}^{u}$ is the $u$-th feature map of $\mathbf{x}_{t+1}$, i.e., $\mathbf{x}_{t+1} = [x_{t+1}^{1}, \ldots, x_{t+1}^{u}, \ldots, x_{t+1}^{C}]$. Here $u$ is the index of output channel. In fact, both Fig.\ref{fig:building_blocks}(c) and Fig.\ref{fig:building_blocks}(e) are special cases of our CB blocks, i.e., Fig.\ref{fig:building_blocks}(f). When
\begin{equation}\label{eqn:exp}
  \mathbf{K}(i, i) = \mathbf{K}(C+i, i) = 1,\ \ \ i = 1, \ldots, C,
\end{equation}
all other elements in $\mathbf{K}$ are 0, and $\lambda_{1} = \lambda_{2} = 1$, then a CB block degrades to the residual block of EDSR \cite{Lim2017Enhanced}. If $\lambda_{1}$ and $\lambda_{2}$ act as learnable weighting factors at this time, then it degrades to the residual block in AWSRN \cite{Wang2019Lightweight}.

In addition, our WCC can also achieve attention mechanism \cite{Hu2017Squeeze,Zhang2018Image} that can be viewed as a guidance to bias the allocation of available processing resources towards the most informative components of an input \cite{Hu2017Squeeze}. But unlike previous self-attention, 1$\times$1 conv layer contributes to the \textit{joint} attention to the identity mapping and the output features of the nonlinear mapping branch.

\begin{figure}[t]
  \centering
  \includegraphics[width=0.99\textwidth]{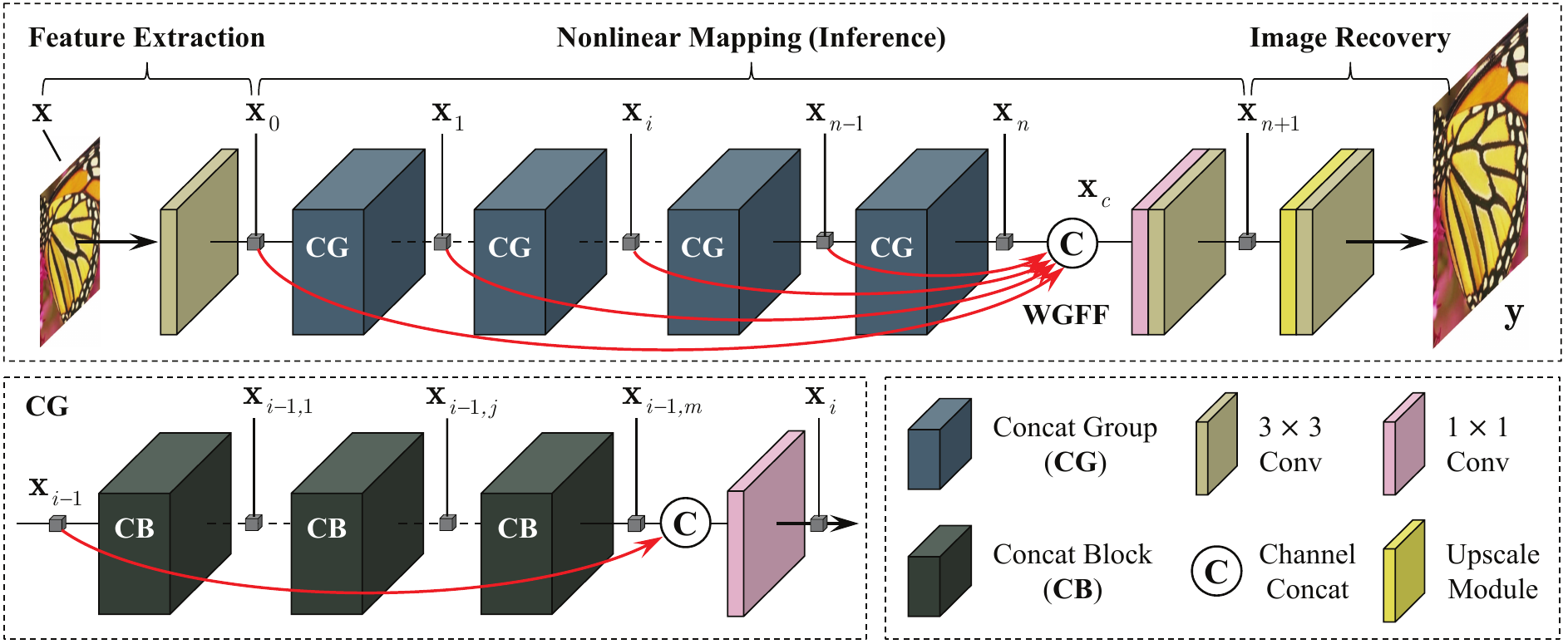}
  \vspace{-2mm}
  \caption{Overall structure of our FC$^2$N model. The {\textcolor[rgb]{1,0,0}{red}} arrows in WGFF and CG denote WCC. Note that there are no residual connections in the entire network.}
\label{fig:FC2N}
\end{figure}

\vspace{-2mm}
\subsection{Overall Network Structure}
\vspace{-1mm}
The overall structure of the FC$^2$N network is shown in Fig.\ref{fig:FC2N}. Similar to many previous models, it mainly includes 3 stages: shallow feature extraction, nonlinear mapping and image reconstruction, which are denoted as $\mathcal{F}_{E}(\cdot)$, $\mathcal{F}_{N}(\cdot)$ and $\mathcal{F}_{R}(\cdot)$ respectively. Assume that the model takes $\mathbf{x}$ as input and outputs $\mathbf{y}$. The shallow features are first extracted by a single 3$\times$3 conv layer:
\begin{equation}\label{eqn:ext}
  \mathbf{x}_{0} = \mathcal{F}_{E}(\mathbf{x}),
\end{equation}
where $\mathbf{x}_{0}$ is the extracted shallow feature maps, and $\mathcal{F}_{E}(\cdot)$ denotes the 3$\times$3 conv layer. Subsequently, $\mathbf{x}_{0}$ is fed into the nonlinear mapping subnet for nonlinear inference. This generates deep features and can be formulated as:
\begin{equation}\label{eqn:nmn}
  \mathbf{x}_{n+1} = \mathcal{F}_{N}(\mathbf{x}_{0}),
\end{equation}
where $\mathbf{x}_{n+1}$ is the generated features, and $\mathcal{F}_{N}(\cdot)$ represents the entire nonlinear mapping, which consists of $n$ cascaded CGs combined with the WGFF. Let's denote the function of the $i$-th CG as $\mathcal{G}_{i}(\cdot)$, i.e., $\mathbf{x}_{i} = \mathcal{G}_{i}(\mathbf{x}_{i-1})$.
Then we can obtain the nonlinear mapping from $\mathbf{x}_{0}$ to $\mathbf{x}_{n}$ iteratively:
\begin{equation}\label{eqn:nonlinear}
  \mathbf{x}_{n} = \mathcal{G}_{n}(\mathbf{x}_{n-1}) =  \mathcal{G}_{n}\Big[ \mathcal{G}_{n-1}\Big(\cdots \mathcal{G}_{1}(\mathbf{x}_{0})\cdots\Big)\Big].
\end{equation}
It is worth noting that simply stacking multiple CGs can easily lead to training failure. Therefore, we use WGFF to integrate intermediate features at different depths, which also helps ease the training difficulty. The deep features $\mathbf{x}_{n+1}$ is therefore obtained by:
\begin{equation}\label{eqn:deep-features}
  \mathbf{x}_{n+1} = \mathcal{F}_{D}(\mathbf{x}_{c}) = \mathcal{F}_{D}([{\lambda}_{0}\mathbf{x}_{0}, {\lambda}_{1}\mathbf{x}_{1}, \ldots, {\lambda}_{n}\mathbf{x}_{n}]),
\end{equation}
where $\lambda_{i}$ is the weighting factor for the corresponding intermediate feature. $\mathcal{F}_{D}(\cdot)$ corresponds to the $1\times1$ conv followed by a $3\times3$ conv, and $[\ldots]$ represents the channel concatenation. This WGFF can further explore model representational capacity by adding negligible model parameters. As in \cite{Lim2017Enhanced,Zhang2018Residual,Zhao2019Channel}, we adopt the ESPCNN \cite{Shi2016Real} to upsample deep features and a 3$\times$3 conv layer to recover HR images: $\hat{\mathbf{y}} = \mathcal{F}_{R}(\mathbf{x}_{n+1})$, where $\hat{\mathbf{y}}$ is the final SR output and $\mathcal{F}_{R}(\cdot)$ corresponds to the upscale module followed by a $3\times3$ conv layer.

We choose to optimize $L_1$ loss as previous works \cite{Lim2017Enhanced,Zhang2018Image,Zhao2019Channel}. Given a training set $\{\mathbf{x}^{(i)}, \mathbf{y}^{(i)}\}_{i = 1}^{N}$, where $N$ is the total number of training samples, the $L_1$ loss is given by:
\begin{equation}\label{eqn:loss}
  L_1(\boldsymbol{\theta}) = \frac{1}{N}\sum_{i=1}^{N}\big\Vert\mathcal{F}\big(\mathbf{x}^{(i)}; \boldsymbol{\theta}\big) - \mathbf{y}^{(i)}\big\Vert_{1},
\end{equation}
where $\mathcal{F}(\cdot)$ corresponds to the entire network, and $\boldsymbol{\theta}$ denotes the parameter set of the model. This is optimized by the Adam \cite{Kingma2014Adam} method. More details on model training will be shown in section \ref{subsec:settings}.

\begin{table}[t]
  \centering
  \small
  \caption{Quantitative results of ablation study on the proposed WCC. All models are evaluated on Set5 \cite{Bevilacqua2012Low} with SR$\times$4, and valid curves are shown in Fig.\ref{fig:AI-compare} ($m = 8, n = 16$)}
  \vspace{-2mm}
  \hspace{-2mm}
  \resizebox{\textwidth}{!}{
  \begin{tabular}{C{1.3cm}|C{1.6cm}|C{1.0cm}|C{1.0cm}|C{1.0cm}|C{1.0cm}|C{1.0cm}|C{1.0cm}|C{1.0cm}|C{1.0cm}|C{1.0cm}}
    \toprule
    \multirow{3}{*}{Configs}& CB   & / & 0 & 1 & 0 & 1 & 0 & 1 & 0 & 1 \\
    \cmidrule{2-11}
                            & CG   & / & 0 & 0 & 1 & 1 & 0 & 0 & 1 & 1 \\
    \cmidrule{2-11}
                            & WGFF & / & 0 & 0 & 0 & 0 & 1 & 1 & 1 & 1 \\
    \midrule
    SR$\times$4             & PSNR & 32.53 & 32.62 & 32.63 & 32.59 & 32.64 & 32.62 & 32.62 & {\textcolor[rgb]{0,0,1}{32.65}} & {\textcolor[rgb]{1,0,0}{32.67}} \\
    \bottomrule
  \end{tabular}}
  \label{tab:ablation_study}
\end{table}

\vspace{-3mm}
\subsection{Concat Group (Concat In Concat)}
\vspace{-2mm}
A CG is simply composed of $m$ stacked CBs with an additional WCC, as shown in Fig.\ref{fig:FC2N}. Let $\mathbf{x}_{i-1} = \mathbf{x}_{i-1, 0}$ be the input of the first CB in the $i$-th CG, then we can obtain the local features as following:
\begin{equation}\label{eqn:CB}
  \mathbf{x}_{i-1,j} = \mathcal{B}_{i, j}(\mathbf{x}_{i-1, j-1}), \ \ \ \ j = 1, 2, \ldots, m,
\end{equation}
where $\mathcal{B}_{i, j}(\cdot)$ represents the function corresponding to the $j$-th CB in the $i$-th CG. Similar to the entire nonlinear mapping, we have:
\begin{equation}\label{eqn:CBm}
  \mathbf{x}_{i-1, m} = \mathcal{B}_{i, m}\Big[\mathcal{B}_{i, m-1}\Big(\cdots \mathcal{B}_{i, 1}(\mathbf{x}_{i-1, 0}) \cdots\Big)\Big].
\end{equation}
To promote the information flow of the network, we further adopt WCC to merge the input of the first CB and the output of the last CB, as shown in Fig.\ref{fig:FC2N}. This CIC structure can not only ease local representation learning, but also allow the linear and nonlinear features are fused in a fine-grained manner. Therefore, we generate the final output of the $i$-th CG:
\begin{equation}\label{eqn:CB-out}
  \mathbf{x}_{i} = \mathcal{L}_{i}([\lambda_{i-1, 0}\mathbf{x}_{i-1, 0},\ \lambda_{i-1, m}\mathbf{x}_{i-1, m}]),
\end{equation}
where $\mathcal{L}_{i}(\cdot)$ stands for the $1\times1$ conv in the $i$-th CG, and $[\ldots]$ denotes channel concatenation. $\lambda_{i-1, 0}$ and $\lambda_{i-1, m}$ are two weighting factors. These operations are similar to local feature fusion (LFF) in some previous work, such as \cite{Zhang2018Residual} and \cite{Zhang2018Image}. However, we use WCC followed by a $1\times1$ conv layer instead of a $3\times3$ conv layer followed by a residual connection \cite{Zhang2018Image} to fuse local features.

\vspace{-2mm}
\subsection{Implementation Details}
\vspace{-1mm}
\label{subsec:implement}
To verify the effective mining of model representational capacity, we implement both largescale and lightweight FC$^2$N models by setting $m = 8, n = 16$ and $m = 4, n = 4$ respectively. In addition, we also use \textit{wide activation} \cite{Yu2018Wide} strategy to allow more low-level information pass through the network while still keep its highly nonlinearity. Specifically, the number of feature channels in the CB is set to $\{32, 128, 32\}$. By shrinking the input and output features and extending the features before ReLU layers, wide activation favors to activating more low-level information without increasing model parameters \cite{Yu2018Wide,Wang2019Lightweight}.

The last layer has 3 filters as it outputs RGB images. Elsewhere, the number of feature channels is set to 32. Except for the 1$\times$1 conv layers annotated in Fig.\ref{fig:building_blocks} and Fig.\ref{fig:FC2N}, all other layers have $3\times3$ kernels, and zero-padding is applied to keep the spatial size of features unchanged. For interlayer connections in WGFF, CG and CB, all weighting factors are learnable and initialized as 1.0.

\vspace{-3mm}
\section{Experimental Results}
\vspace{-2mm}
\subsection{Settings}
\vspace{-1mm}
\label{subsec:settings}
As in \cite{Lim2017Enhanced,Zhang2018Residual,Zhang2018Image,Haris2018Deep}, we adopt 800 training images from DIV2K dataset \cite{Agustsson2017NTIRE} as our training set. Data augmentation is performed on training images by randomly horizontal and vertical flips, $90^{\circ}$ rotations and data range complementarity. Five benchmark datasets, including Set5 \cite{Bevilacqua2012Low}, Set14 \cite{Zeyde2010On}, B100 \cite{Martin2001A}, Urban100 \cite{Huang2015Single} and Manga109 \cite{Matsui2017Sketch}, are used for evaluation. The results are typically evaluated with PSNR and SSIM \cite{Wang2004Image} on Y channel of YCbCr space. For model training, $48\times48$ LR image patches are extracted from LR images, while the size of HR patches corresponds to the scaling factors. Batch size is set to 16 as in previous works \cite{Lim2017Enhanced,Zhang2018Residual,Zhang2018Image}. The objective function is minimized by the Adam optimizer \cite{Kingma2014Adam} with $\beta_{1} = 0.9$, $\beta_{2} = 0.999$ and $\epsilon = 10^{-8}$. The learning rate is initialized as $2\times10^{-4}$ for all layers and halved for every $4.0 \times 10^5$ training steps. Both largescale and lightweight FC$^2$N models are trained for $10^6$ iterations in total. To compare the computational overhead of deep models in the lightweight case, we also introduce MultiAdds \cite{Ahn2018Fast} as an evaluation metric\footnote[2]{The source code of our FC$^2$N is available at \url{https://github.com/zxlation/FC2N}.}.

\begin{figure}[t]
  \centering
  \begin{minipage}[c]{0.49\textwidth}
    \centering
    \includegraphics[width=0.9\textwidth]{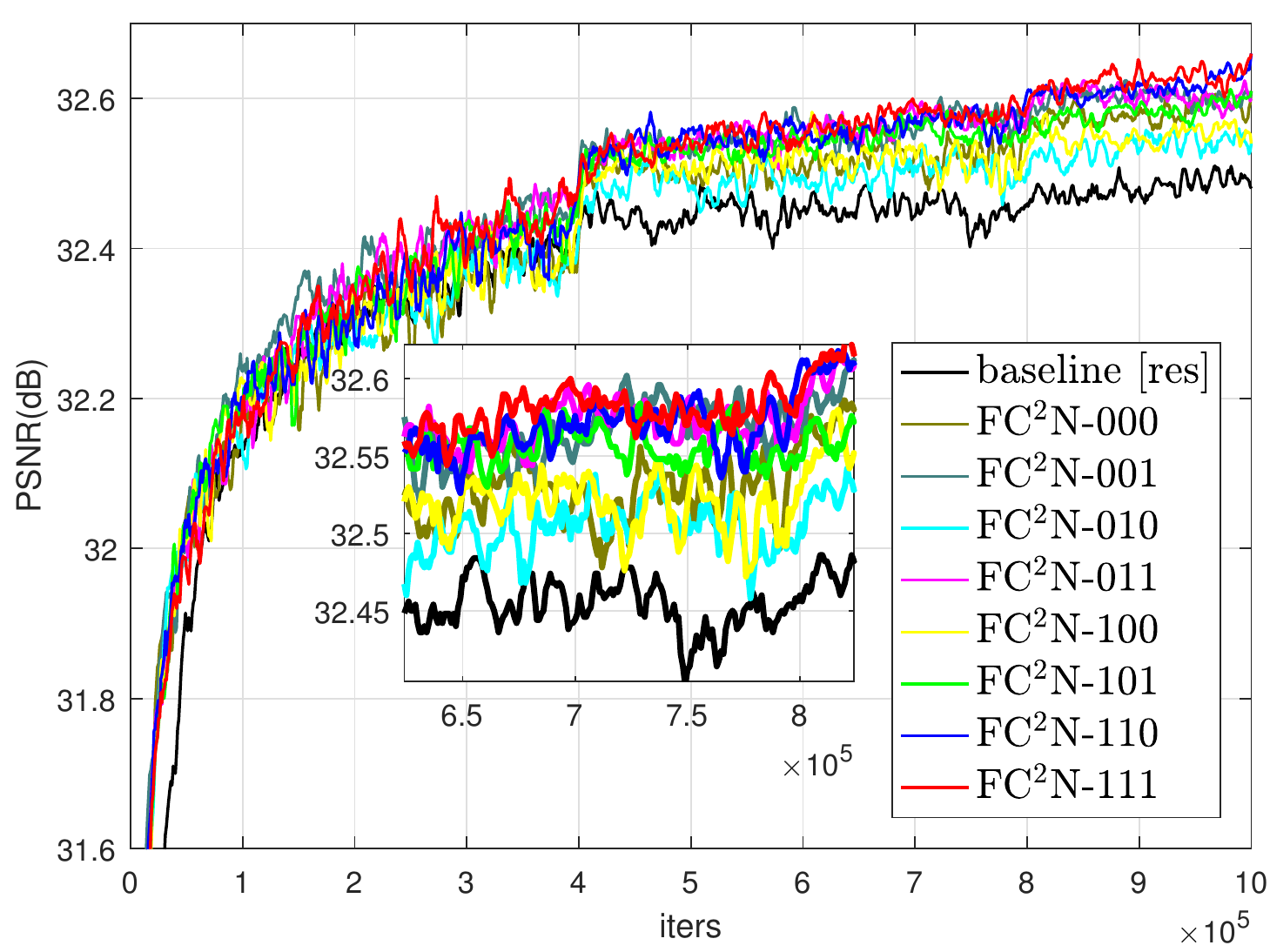}
    \vspace{-4mm}
    \caption{Valid PSNR curves on the WCC}
    \label{fig:AI-compare}
  \end{minipage}%
  \hspace{1mm}
  \begin{minipage}[c]{0.49\textwidth}
    \centering
    \includegraphics[width=0.9\textwidth]{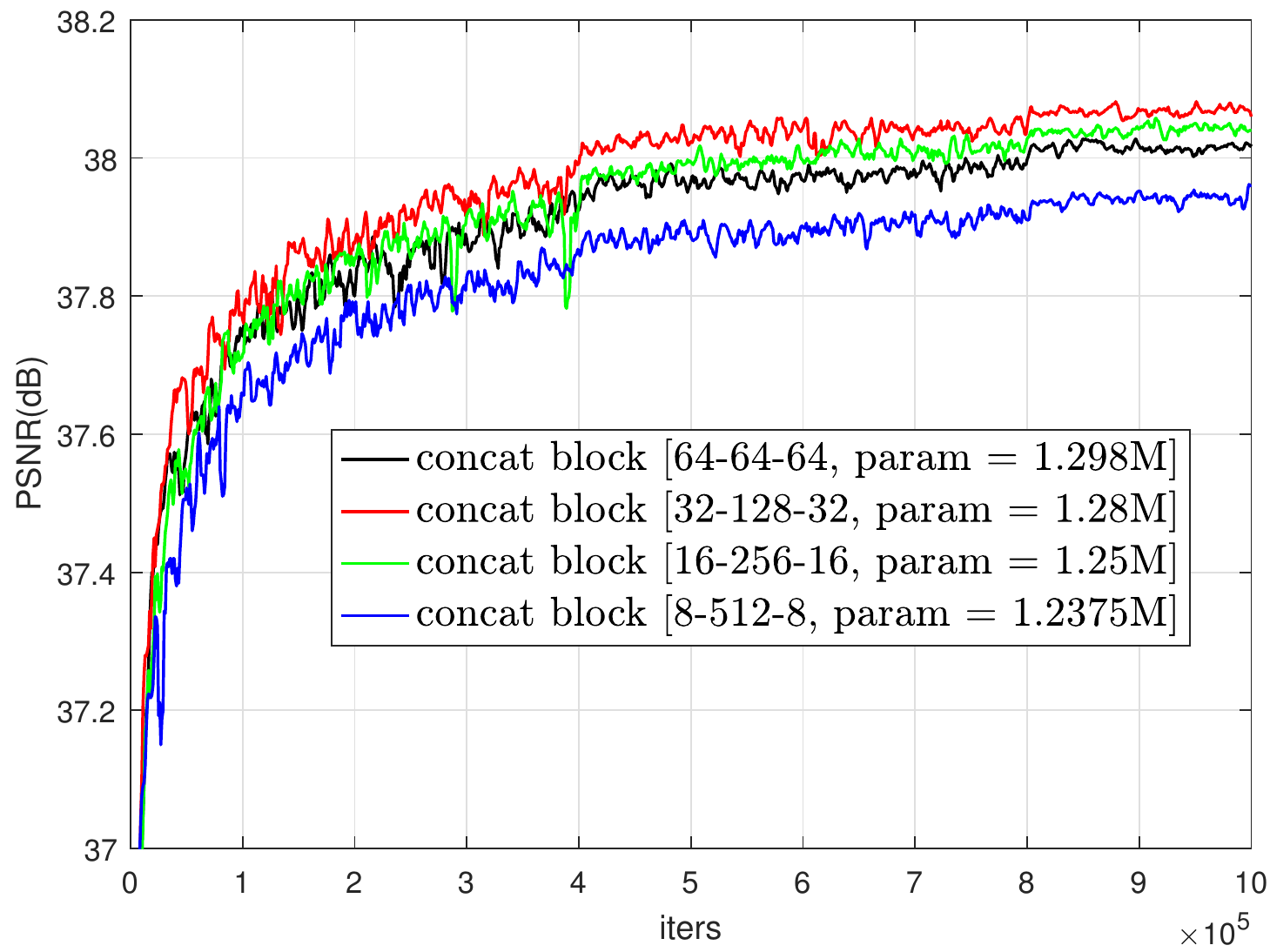}
    \vspace{-4mm}
    \caption{Valid curves on wide activation}
    \label{fig:wide-activation}
  \end{minipage}
\end{figure}

\begin{table}[t]
  \centering
  \footnotesize
  \caption{Impact of $n$ and $m$ on model performance (PSNR / params)}
  \vspace{-2mm}
  \begin{tabular}{C{1.0cm}|C{2.6cm}|C{2.6cm}|C{2.6cm}|C{2.6cm}}
    \toprule
    $n \backslash m$ & 2 & 4 & 6 & 8 \\
    \hline
    2  & 31.75dB / 0.40M & 32.03dB / 0.70M & 32.12dB / 1.01M & 32.31dB / 1.31M \\
    \hline
    4  & 32.02dB / 0.71M & 32.23dB / 1.31M & 32.33dB / 1.92M & 32.40dB / 2.53M \\
    \hline
    8  & 32.23dB / 1.33M & 32.36dB / 2.54M & 32.52dB / 3.76M & 32.56dB / 4.97M \\
    \hline
    16 & 32.40dB / 2.57M & 32.53dB / 5.00M & 32.62dB / 7.43M & 32.67dB / 9.86M \\
    \bottomrule
  \end{tabular}
  \label{tab:CG-CB}
\end{table}

\begin{figure}[t]
  \centering
  \includegraphics[width=0.92\textwidth]{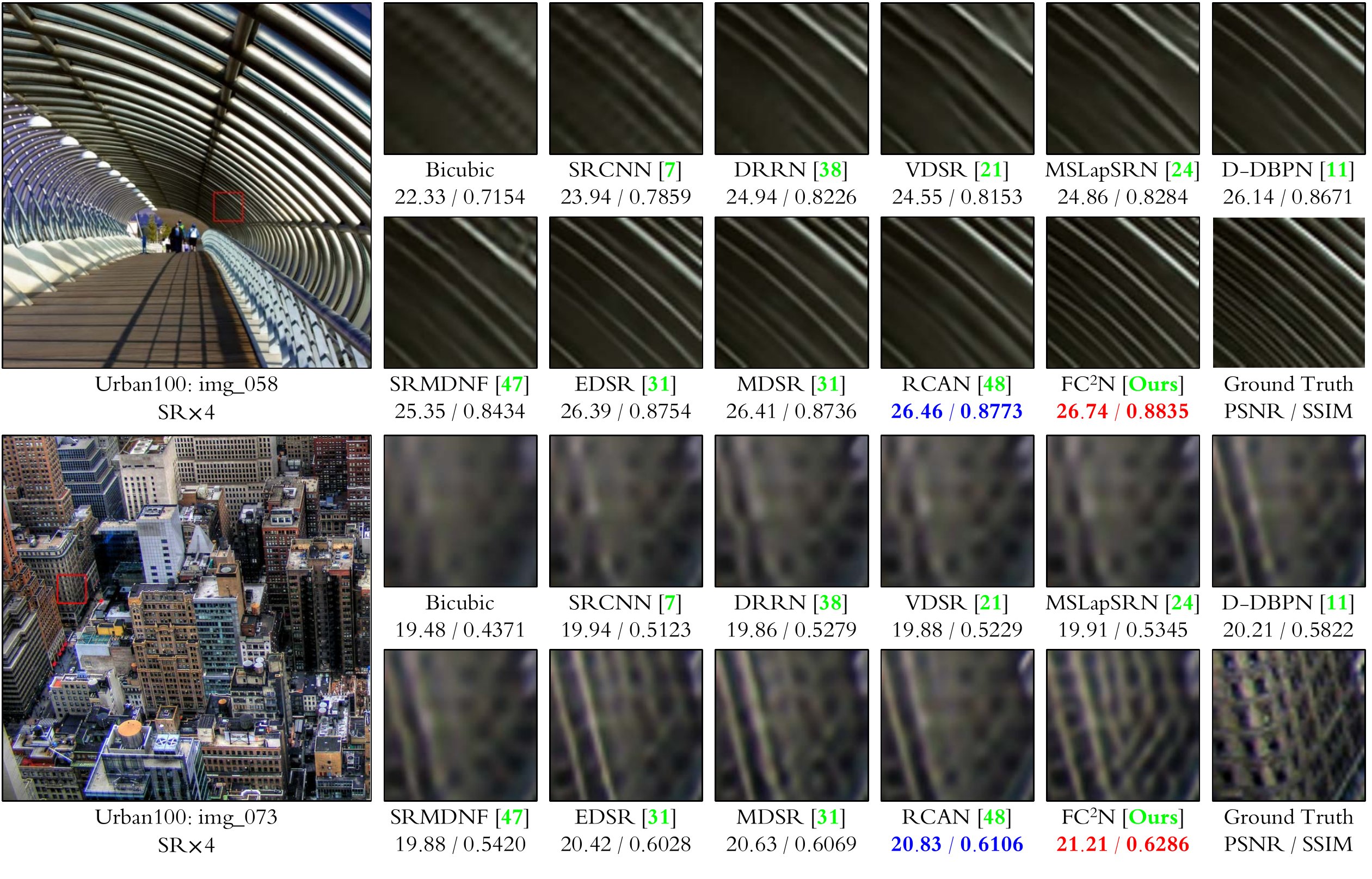}
  \vspace{-3mm}
  \caption{Visual comparison between other SR methods and our largescale FC$^2$N. The best and second best results are marked in {\textcolor[rgb]{1,0,0}{red}} and {\textcolor[rgb]{0,0,1}{blue}} respectively.}
  \label{fig:vis_largescale_FC2N}
\end{figure}

\vspace{-3mm}
\subsection{Model Analysis}
\vspace{-1mm}
\subsubsection{Weighted Channel Concatenation}
To show the superiority of our WCC to residual learning, we build a baseline residual model, in which all skip connections in CGs and CBs are replaced by \textit{unweighted} residual connections and WGFF is changed to GFF \cite{Zhang2018Residual}. This model corresponds to the ``baseline [res]'' in Fig.\ref{fig:AI-compare} and the first column in Table \ref{tab:ablation_study} (PSNR = 32.53 dB). To investigate the impact of learnable weighting factors, we also conduct ablation study on whether channel concatenations in CB, CG and WGFF are weighted. They are indicated by ``1'' if they are weighted by learnable parameters, otherwise denoted as ``0''. Notations and quantitative results are shown in Fig.\ref{fig:AI-compare} and Table \ref{tab:ablation_study}. Note that three numbers representing model suffix correspond to WGFF, CG and CB respectively. For example, FC$^2$N-110 represents the model with WGFF and CG are weighted but CB is unweighted, which corresponds to the penultimate column of Table \ref{tab:ablation_study}. In addition, we find that eliminating WCC in either CB, CG or WGFF is likely to result in training failure, so we do not perform ablation study in this case.

\textit{Comparison with Residual Baseline}: As illustrated in Fig.\ref{fig:AI-compare} and Table \ref{tab:ablation_study}, by comparing with FC$^2$N-000, it can be seen that about 0.1dB performance gain can be achieved by changing the residual connections in CGs and CBs to \textit{unweighted} channel concatenations. The validation curves of baseline [res] and FC$^2$N-000 in Fig.\ref{fig:AI-compare} also illustrate that, after the decays of learning rate, channel concatenation can make better SR performance improvement than residual connection although both models have roughly the same model parameters (9.6M vs. 9.8M).

\textit{Learnable Weighting Factors}: In Table \ref{tab:ablation_study}, by comparing FC$^2$N-001, FC$^2$N-010 and FC$^2$N-100 with FC$^2$N-000, we can see that learnable weighting factors in all WGFF, CGs and CBs favor to performance improvement. In addition, the best performance given by FC$^2$N-111 shows that the WCC in WGFF, CGs and CBs is beneficial to maximizing the performance gain. In particular, WGFF regularly integrates shallow and deep features in an adaptive manner, hence improving the information flow throughout the entire network. This can also be verified by the comparison between FC$^2$N-011 and FC$^2$N-111. In fact, the proposed WCC also helps stabilize model training. Although Table \ref{tab:ablation_study} presents the results of all configurations, the models that do not make adequate use of weighting, such as FC$^2$N-000, FC$^2$N-001 and FC$^2$N-010 etc., show some training difficulty as they crashed many times during training. The convergence curves shown in Fig.\ref{fig:AI-compare} also verify the above analysis.

\begin{table}[t]
  \centering
  \caption{Quantitative comparison w.r.t \textbf{large-scale} FC$^2$N ($n = 16, m = 8$). Best and second best results are marked in {\textcolor[rgb]{1,0,0}{red}} and {\textcolor[rgb]{0,0,1}{blue}} respectively (PSNR (dB)/SSIM).}
  \vspace{-2mm}
  \hspace{-2mm}
  \resizebox{\textwidth}{!}{
  \begin{tabular}{R{2.4cm}|C{0.6cm}|C{1.2cm}|C{2.4cm}|C{2.4cm}|C{2.4cm}|C{2.4cm}|C{2.4cm}}
    \toprule
    Methods & $r$ &  Params  & Set5 \cite{Bevilacqua2012Low} & Set14 \cite{Zeyde2010On} & B100 \cite{Martin2001A} & Urban100 \cite{Huang2015Single} & Manga109 \cite{Matsui2017Sketch} \\
    \hline
    Bicubic                         &  2  &   /   & 33.66 / 0.9299 & 30.24 / 0.8688 & 29.56 / 0.8431 & 26.88 / 0.8403 & 30.80 / 0.9339 \\
    EDSR \cite{Lim2017Enhanced}     &  2  & 40.7M & 38.11 / 0.9602 & 33.92 / 0.9195 & 32.32 / 0.9013 & 32.93 / 0.9351 & 39.10 / 0.9773 \\
    D-DBPN \cite{Haris2018Deep}     &  2  & 5.82M & 38.09 / 0.9600 & 33.85 / 0.9190 & 32.27 / 0.9000 & 32.55 / 0.9324 & 38.89 / 0.9775 \\
    RDN \cite{Zhang2018Residual}    &  2  & 22.1M & 38.24 / 0.9614 & 34.01 / 0.9212 & 32.34 / 0.9017 & 32.89 / 0.9353 & 39.18 / 0.9780 \\
    RCAN \cite{Zhang2018Image}      &  2  & 15.4M & 38.27 / 0.9614 & 34.11 / 0.9216 & 32.41 / 0.9026 & 33.34 / 0.9384 & 39.44 / 0.9786 \\
    SAN \cite{Dai2019Second}        &  2  & 15.7M & 38.31 / 0.9620 & 34.07 / 0.9213 & 32.42 / 0.9028 & 33.10 / 0.9370 & 39.32 / 0.9790 \\
    SRFBN \cite{Li2019Feedback}     &  2  &  N/A  & 38.11 / 0.9609 & 33.82 / 0.9196 & 32.29 / 0.9010 & 33.62 / 0.9328 & 39.08 / 0.9779 \\
    OISR \cite{He2019ODE}           &  2  &  N/A  & 38.21 / 0.9612 & 33.94 / 0.9206 & 32.36 / 0.9019 & 33.03 / 0.9365 & N/A \\
    FC$^2$N        [Ours]           &  2  & 9.82M & 38.29 / 0.9616 & 34.14 / 0.9224 & 32.40 / 0.9025 & 33.18 / 0.9379 & 39.51 / 0.9787 \\
    FC$^2$N$^{+}$  [Ours]           &  2  & 9.82M & {\textcolor[rgb]{0,0,1}{38.33 / 0.9617}} & {\textcolor[rgb]{0,0,1}{34.24 / 0.9224}} & {\textcolor[rgb]{0,0,1}{32.44 / 0.9029}} & {\textcolor[rgb]{0,0,1}{33.34 / 0.9388}} & {\textcolor[rgb]{0,0,1}{39.62 / 0.9790}} \\
    FC$^2$N$^{++}$ [Ours]           &  2  & 9.82M & {\textcolor[rgb]{1,0,0}{38.34 / 0.9618}} & {\textcolor[rgb]{1,0,0}{34.25 / 0.9225}} & {\textcolor[rgb]{1,0,0}{32.45 / 0.9030}} & {\textcolor[rgb]{1,0,0}{33.39 / 0.9392}} & {\textcolor[rgb]{1,0,0}{39.68 / 0.9792}} \\
    \hline
    Bicubic                         &  3  &   /   & 30.39 / 0.8682 & 27.55 / 0.7742 & 27.21 / 0.7385 & 24.46 / 0.7349 & 26.95 / 0.8556 \\
    EDSR \cite{Lim2017Enhanced}     &  3  & 43.7M & 34.65 / 0.9280 & 30.52 / 0.8462 & 29.25 / 0.8093 & 28.80 / 0.8653 & 34.17 / 0.9476 \\
    RDN \cite{Zhang2018Residual}    &  3  & 22.3M & 34.71 / 0.9296 & 30.57 / 0.8468 & 29.26 / 0.8093 & 28.80 / 0.8653 & 34.13 / 0.9484 \\
    RCAN \cite{Zhang2018Image}      &  3  & 15.6M & 34.74 / 0.9299 & 30.65 / 0.8482 & 29.32 / 0.8111 & 29.09 / 0.8702 & 34.44 / 0.9499 \\
    SAN \cite{Dai2019Second}        &  3  & 15.9M & 34.75 / 0.9300 & 30.59 / 0.8476 & 29.33 / 0.8112 & 28.93 / 0.8671 & 34.30 / 0.9494 \\
    SRFBN \cite{Li2019Feedback}     &  3  &  N/A  & 34.70 / 0.9292 & 30.51 / 0.8461 & 29.24 / 0.8084 & 28.73 / 0.8641 & 34.18 / 0.9481 \\
    OISR \cite{He2019ODE}           &  3  &  N/A  & 34.72 / 0.9297 & 30.57 / 0.8470 & 29.29 / 0.8103 & 28.95 / 0.8680 & N/A \\
    FC$^2$N        [Ours]           &  3  & 9.87M & 34.76 / 0.9302 & 30.66 / 0.8485 & 29.31 / 0.8106 & 29.04 / 0.8700 & 34.63 / 0.9504 \\
    FC$^2$N$^{+}$  [Ours]           &  3  & 9.87M & {\textcolor[rgb]{0,0,1}{34.85 / 0.9307}} & {\textcolor[rgb]{0,0,1}{30.76 / 0.8495}} & {\textcolor[rgb]{0,0,1}{29.36 / 0.8114}} & {\textcolor[rgb]{0,0,1}{29.22 / 0.8725}} & {\textcolor[rgb]{0,0,1}{34.87 / 0.9514}} \\
    FC$^2$N$^{++}$ [Ours]           &  3  & 9.87M & {\textcolor[rgb]{1,0,0}{34.85 / 0.9307}} & {\textcolor[rgb]{1,0,0}{30.78 / 0.8497}} & {\textcolor[rgb]{1,0,0}{29.37 / 0.8115}} & {\textcolor[rgb]{1,0,0}{29.24 / 0.8727}} & {\textcolor[rgb]{1,0,0}{34.91 / 0.9515}} \\
    \hline
    Bicubic                         &  4  &   /   & 28.42 / 0.8104 & 26.00 / 0.7027 & 25.96 / 0.6675 & 23.14 / 0.6577 & 24.89 / 0.7866 \\
    EDSR \cite{Lim2017Enhanced}     &  4  & 43.1M & 32.46 / 0.8968 & 28.80 / 0.7876 & 27.71 / 0.7420 & 26.64 / 0.8033 & 31.02 / 0.9148 \\
    D-DBPN \cite{Haris2018Deep}     &  4  & 10.4M & 32.47 / 0.8980 & 28.82 / 0.7860 & 27.72 / 0.7400 & 26.38 / 0.7946 & 30.91 / 0.9137 \\
    RDN \cite{Zhang2018Residual}    &  4  & 22.3M & 32.47 / 0.8990 & 28.81 / 0.7871 & 27.72 / 0.7419 & 26.61 / 0.8028 & 31.00 / 0.9151 \\
    RCAN \cite{Zhang2018Image}      &  4  & 15.6M & 32.63 / 0.9002 & 28.87 / 0.7889 & 27.77 / 0.7436 & 26.82 / 0.8087 & 31.22 / 0.9173 \\
    SAN \cite{Dai2019Second}        &  4  & 15.9M & 32.64 / 0.9003 & 28.92 / 0.7888 & 27.78 / 0.7436 & 26.79 / 0.8068 & 31.18 / 0.9169 \\
    SRFBN \cite{Li2019Feedback}     &  4  &  N/A  & 32.47 / 0.8983 & 28.81 / 0.7868 & 27.72 / 0.7409 & 26.60 / 0.8015 & 31.15 / 0.9160 \\
    OISR \cite{He2019ODE}           &  4  &  N/A  & 32.53 / 0.8992 & 28.86 / 0.7878 & 27.75 / 0.7428 & 26.64 / 0.8033 & N/A \\
    FC$^2$N        [Ours]           &  4  & 9.86M & 32.67 / 0.9005 & 28.90 / 0.7889 & 27.77 / 0.7432 & 26.85 / 0.8089 & 31.41 / 0.9193 \\
    FC$^2$N$^{+}$  [Ours]           &  4  & 9.86M & {\textcolor[rgb]{0,0,1}{32.73 / 0.9011}} & {\textcolor[rgb]{0,0,1}{29.00 / 0.7906}} & {\textcolor[rgb]{0,0,1}{27.83 / 0.7445}} & {\textcolor[rgb]{0,0,1}{27.01 / 0.8127}} & {\textcolor[rgb]{0,0,1}{31.71 / 0.9215}} \\
    FC$^2$N$^{++}$ [Ours]           &  4  & 9.86M & {\textcolor[rgb]{1,0,0}{32.74 / 0.9012}} & {\textcolor[rgb]{1,0,0}{29.02 / 0.7909}} & {\textcolor[rgb]{1,0,0}{27.83 / 0.7446}} & {\textcolor[rgb]{1,0,0}{27.03 / 0.8130}} & {\textcolor[rgb]{1,0,0}{31.74 / 0.9216}} \\
    \hline
    Bicubic                         &  8  &   /   & 24.40 / 0.6580 & 23.10 / 0.5660 & 23.67 / 0.5480 & 20.74 / 0.5160 & 21.47 / 0.6500 \\
    EDSR \cite{Lim2017Enhanced}     &  8  & 45.5M & 26.96 / 0.7762 & 24.91 / 0.6420 & 24.81 / 0.5985 & 22.51 / 0.6221 & 24.69 / 0.7841 \\
    D-DBPN \cite{Haris2018Deep}     &  8  & 23.2M & 27.21 / 0.7840 & 25.13 / 0.6480 & 24.88 / 0.6010 & 22.73 / 0.6312 & 25.14 / 0.7987 \\
    RCAN \cite{Zhang2018Image}      &  8  & 15.7M & 27.31 / {\textcolor[rgb]{0,0,1}{0.7878}} & 25.23 / 0.6511 & {\textcolor[rgb]{0,0,1}{24.98}} / {\textcolor[rgb]{1,0,0}{0.6058}} & {\textcolor[rgb]{1,0,0}{23.00 / 0.6452}} & 25.24 / {\textcolor[rgb]{1,0,0}{0.8029}} \\
    SAN \cite{Dai2019Second}        &  8  & 16.1M & 27.22 / 0.7829 & 25.14 / 0.6476 & 24.88 / 0.6011 & 22.70 / 0.6314 & 24.85 / 0.7906 \\
    FC$^2$N        [Ours]           &  8  & 9.90M & 27.25 / 0.7833 & 25.10 / 0.6479 & 24.87 / 0.6016 & 22.72 / 0.6331 & 25.00 / 0.7937 \\
    FC$^2$N$^{+}$  [Ours]           &  8  & 9.90M & {\textcolor[rgb]{1,0,0}{27.35 / 0.7880}} & {\textcolor[rgb]{0,0,1}{25.27 / 0.6512}} & 24.96 / 0.6041 & 22.94 / 0.6398 & {\textcolor[rgb]{0,0,1}{25.34}} / 0.8005 \\
    FC$^2$N$^{++}$ [Ours]           &  8  & 9.90M & {\textcolor[rgb]{0,0,1}{27.35}} / 0.7872 & {\textcolor[rgb]{1,0,0}{25.29 / 0.6517}} & {\textcolor[rgb]{1,0,0}{25.01}} / {\textcolor[rgb]{0,0,1}{0.6043}} & {\textcolor[rgb]{0,0,1}{22.97 / 0.6407}} & {\textcolor[rgb]{1,0,0}{25.38}} / {\textcolor[rgb]{0,0,1}{0.8015}} \\
    \bottomrule
  \end{tabular}}
  \label{tab:quantcomp}
\end{table}

\vspace{-4mm}
\subsubsection{Wide Activation}
In the context of single image SR, low-level information may contribute to more accurate pixel-wise predication \cite{Yu2018Wide,Wang2019Lightweight}, and wide activation is considered to help promote the propagation of low-level features in the network and boost model performance. To verify the effectiveness of wide activation, we compare several configurations in this section. Fig.\ref{fig:wide-activation} shows the validation curves of the lightweight FC$^2$N ($n = 4, m = 4$) on Set5 \cite{Bevilacqua2012Low} with SR$\times$2, which is equipped with different configurations of wide activation.

Assume that the width of identity mapping pathway is $w_1$ and that of the nonlinear mapping pathway is $w_2$. Let $r_{wa}$ denote the ratio of $w_1$ and $w_2$:
\begin{equation}\label{eqn:ratio}
  r_{wa} = \frac{w_2}{w_1}.
\end{equation}
For fairness of comparison, we keep the feature width of feature extraction and image reconstruction the same as that of FC$^2$N. It can be seen from Fig.\ref{fig:wide-activation} that properly increasing $r_{wa}$ favors to performance improvement, e.g., $r_{wa} = 1$ and $r_{wa} = 4$. However, model performance will degrade as $r_{wa}$ continues to increase and reaches a certain threshold, e.g., $r_{wa} = 64$. Similar phenomenon was also observed in \cite{Yu2018Wide} and one possible reason for this performance degradation is that the identity mapping becomes too slim, resulting in the bottleneck of low-level information propagation. The observation suggests that the wide activation that is effective in residuals learning is also effective in WCC, which experimentally demonstrates the inclusion of residual connection in WCC.

\vspace{-2mm}
\subsubsection{The Number of CG and CB}
Table \ref{tab:CG-CB} exhibits the testing results of different combinations of $m$ and $n$, on Set5 \cite{Bevilacqua2012Low} with SR$\times$4. It can be seen that the increase in both $m$ and $n$ helps boost the performance of the model, which is unsurprising because increasing $m$ and $n$ obviously enlarges model scale, including network depth and model parameters. It is worth noting that at roughly the same model scale, larger $m$ is more helpful to performance improvement than larger $n$. For instance, the model with $n = 8, m = 2$ performs slightly worse than the model with $n = 2, m = 8$, which has fewer parameters but the same network depth and width. Moreover, we can see that our FC$^2$N achieves excellent SR performance in both lightweight and largescale implementations. This implies that it consistently provides good performance-scale tradeoffs as model scale changes.

\begin{table}[t]
  \centering
  \footnotesize
  \caption{Quantitative comparison with \textbf{lightweight} FC$^2$N ($n = m = 4$). The metric of MultAdds is computed with HR images with size of 720p (1280$\times$720).}
  \vspace{-2mm}
  \hspace{-2mm}
  \resizebox{\textwidth}{!}{
  \begin{tabular}{R{2.4cm}|C{0.6cm}|C{1.8cm}|C{1.8cm}|C{2.40cm}|C{2.40cm}|C{2.40cm}|C{2.40cm}}
    \toprule
    Methods & $r$ & Params & MultiAdds & Set5 \cite{Bevilacqua2012Low} & Set14 \cite{Zeyde2010On} & B100 \cite{Martin2001A} & Manga109 \cite{Matsui2017Sketch} \\
    \hline 
    Bicubic                            & 2 &    /   &    /     & 33.66 / 0.9299 & 30.24 / 0.8688 & 29.56 / 0.8431 & 30.80 / 0.9339  \\
    SRCNN \cite{Dong2016Image}         & 2 &   57K  &  52.7G   & 36.66 / 0.9542 & 32.42 / 0.9063 & 31.36 / 0.8879 & 35.74 / 0.9661  \\
    VDSR \cite{Kim2016Accurate}        & 2 &  665K  &  612.6G  & 37.53 / 0.9587 & 33.03 / 0.9124 & 31.90 / 0.8960 & 37.22 / 0.9729  \\
    DRRN \cite{Tai2017Image}           & 2 &  297K  & 6,796.9G & 37.74 / 0.9591 & 33.23 / 0.9136 & 32.05 / 0.8973 & 37.92 / 0.9760  \\
    LapSRN \cite{Lai2017Deep}          & 2 &  813K  &  29.9G   & 37.52 / 0.9590 & 33.08 / 0.9130 & 31.80 / 0.8950 & 37.27 / 0.9740  \\
    SRMDNF \cite{Zhang2018Learning}    & 2 & 1,513K &  347.7G  & 37.79 / 0.9600 & 33.32 / 0.9150 & 32.05 / 0.8980 & 38.07 / 0.9761  \\
    NLRN \cite{Liu2018NonLocal}        & 2 &  N/A   &   N/A    & 38.00 / 0.9603 & 33.46 / 0.9159 & 32.19 / 0.8992 & N/A \\
    AWSRN \cite{Wang2019Lightweight}   & 2 & 1,397K &  320.5G  & 38.11 / 0.9608 & 33.78 / 0.9189 & {\textcolor[rgb]{0,0,1}{32.26 / 0.9006}} & 38.87 / 0.9776  \\
    CARN \cite{Ahn2018Fast}            & 2 & 1,592K &  222.8G  & 37.76 / 0.9590 & 33.52 / 0.9166 & 32.09 / 0.8978 & N/A \\
    MSRN \cite{Li2018Multi}            & 2 & 5,930K & 1,365.4G & 38.08 / 0.9607 & 33.70 / 0.9186 & 32.23 / 0.9002 & 38.69 / 0.9772  \\
    FC$^2$N        [Ours]              & 2 & 1,277K &  294.0G  & 38.11 / 0.9608 & 33.70 / 0.9179 & 32.21 / 0.9001 & 38.81 / 0.9775 \\
    FC$^2$N$^{+}$  [Ours]              & 2 & 1,277K &  294.0G  & {\textcolor[rgb]{0,0,1}{38.16 / 0.9610}} & {\textcolor[rgb]{0,0,1}{33.78 / 0.9189}} & 32.25 / 0.9006 & {\textcolor[rgb]{0,0,1}{39.02 / 0.9779}} \\
    FC$^2$N$^{++}$ [Ours]              & 2 & 1,277K &  294.0G  & {\textcolor[rgb]{1,0,0}{38.17 / 0.9611}} & {\textcolor[rgb]{1,0,0}{33.79 / 0.9191}} & {\textcolor[rgb]{1,0,0}{32.28 / 0.9007}} & {\textcolor[rgb]{1,0,0}{39.04 / 0.9780}} \\
    \hline
    Bicubic                            & 3 &   /    &    /     & 30.39 / 0.8682 & 27.55 / 0.7742 & 27.21 / 0.7385 & 26.95 / 0.8556  \\
    SRCNN \cite{Dong2016Image}         & 3 &  57K   &  52.7G   & 32.75 / 0.9090 & 29.28 / 0.8209 & 28.41 / 0.7863 & 30.59 / 0.9107  \\
    VDSR \cite{Kim2016Accurate}        & 3 &  665K  &  612.6G  & 33.66 / 0.9213 & 29.77 / 0.8314 & 28.82 / 0.7976 & 32.01 / 0.9310  \\
    DRRN \cite{Tai2017Image}           & 3 &  297K  & 6,796.9G & 34.03 / 0.9244 & 29.96 / 0.8349 & 28.95 / 0.8004 & 32.74 / 0.9390  \\
    SRMDNF \cite{Zhang2018Learning}    & 3 & 1,530K &  156.3G  & 34.12 / 0.9250 & 30.04 / 0.8370 & 28.97 / 0.8030 & 33.00 / 0.9403  \\
    NLRN \cite{Liu2018NonLocal}        & 3 &  N/A   &    N/A   & 34.27 / 0.9266 & 30.16 / 0.8374 & 29.06 / 0.8026 & N/A \\
    AWSRN \cite{Wang2019Lightweight}   & 3 & 1,476K &  150.6G  & 34.52 / 0.9281 & 30.38 / 0.8426 & 29.16 / 0.8069 & 33.85 / 0.9463  \\
    CARN \cite{Ahn2018Fast}            & 3 & 1,592K &  118.8G  & 34.29 / 0.9255 & 30.29 / 0.8407 & 29.06 / 0.8034 & N/A \\
    MSRN \cite{Li2018Multi}            & 3 & 6,114K &  625.7G  & 34.46 / 0.9278 & 30.41 / 0.8437 & 29.15 / 0.8064 & 33.67 / 0.9456  \\
    FC$^2$N        [Ours]              & 3 & 1,323K &  135.8G  & 34.53 / 0.9282 & 30.44 / 0.8437 & 29.16 / 0.8068 & 33.83 / 0.9462 \\
    FC$^2$N$^{+}$  [Ours]              & 3 & 1,323K &  135.8G  & {\textcolor[rgb]{0,0,1}{34.60 / 0.9287}} & {\textcolor[rgb]{0,0,1}{30.51 / 0.8449}} & {\textcolor[rgb]{0,0,1}{29.20 / 0.8074}} & {\textcolor[rgb]{0,0,1}{34.11 / 0.9476}} \\
    FC$^2$N$^{++}$ [Ours]              & 3 & 1,323K &  135.8G  & {\textcolor[rgb]{1,0,0}{34.61 / 0.9287}} & {\textcolor[rgb]{1,0,0}{30.52 / 0.8450}} & {\textcolor[rgb]{1,0,0}{29.21 / 0.8075}} & {\textcolor[rgb]{1,0,0}{34.16 / 0.9478}} \\
    \hline
    Bicubic                            & 4 &   /    &     /    & 28.42 / 0.8104 & 26.00 / 0.7027 & 25.96 / 0.6675 & 24.89 / 0.7866  \\
    SRCNN \cite{Dong2016Image}         & 4 &  57K   &  52.7G   & 30.48 / 0.8628 & 27.49 / 0.7503 & 26.90 / 0.7101 & 27.66 / 0.8505  \\
    VDSR \cite{Kim2016Accurate}        & 4 &  665K  &  612.6G  & 31.35 / 0.8838 & 28.01 / 0.7674 & 27.29 / 0.7251 & 28.83 / 0.8809  \\
    DRRN \cite{Tai2017Image}           & 4 &  297K  & 6,796.9G & 31.68 / 0.8888 & 28.21 / 0.7720 & 27.38 / 0.7284 & 29.46 / 0.8960  \\
    LapSRN \cite{Lai2017Deep}          & 4 &  813K  & 149.4G   & 31.54 / 0.8850 & 28.19 / 0.7720 & 27.32 / 0.7280 & 29.09 / 0.8845  \\
    SRMDNF \cite{Zhang2018Learning}    & 4 & 1,555K &   89.3G  & 31.96 / 0.8930 & 28.35 / 0.7770 & 27.49 / 0.7340 & 30.09 / 0.9024  \\
    NLRN \cite{Liu2018NonLocal}        & 4 &  N/A   &   N/A    & 31.92 / 0.8916 & 28.36 / 0.7745 & 27.48 / 0.7346 & N/A \\
    AWSRN \cite{Wang2019Lightweight}   & 4 & 1,587K &   91.1G  & 32.27 / 0.8960 & 28.69 / 0.7843 & 27.64 / 0.7385 & 30.72 / 0.9109 \\
    CARN \cite{Ahn2018Fast}            & 4 & 1,592K &   90.9G  & 32.13 / 0.8937 & 28.60 / 0.7806 & 27.58 / 0.7349 & N/A \\
    MSRN \cite{Li2018Multi}            & 4 & 6,078K &  349.8G  & 32.26 / 0.8960 & 28.63 / 0.7836 & 27.61 / 0.7380 & 30.57 / 0.9103 \\
    FC$^2$N        [Ours]              & 4 & 1,314K &   82.6G  & 32.23 / 0.8956 & 28.68 / 0.7836 & 27.62 / 0.7377 & 30.74 / 0.9110 \\
    FC$^2$N$^{+}$  [Ours]              & 4 & 1,314K &   82.6G  & {\textcolor[rgb]{0,0,1}{32.36 / 0.8970}} & {\textcolor[rgb]{0,0,1}{28.75 / 0.7851}} & {\textcolor[rgb]{0,0,1}{27.68 / 0.7390}} & {\textcolor[rgb]{0,0,1}{31.04 / 0.9136}} \\
    FC$^2$N$^{++}$ [Ours]              & 4 & 1,314K &   82.6G  & {\textcolor[rgb]{1,0,0}{32.37 / 0.8971}} & {\textcolor[rgb]{1,0,0}{28.76 / 0.7853}} & {\textcolor[rgb]{1,0,0}{27.68 / 0.7391}} & {\textcolor[rgb]{1,0,0}{31.06 / 0.9137}} \\
    \bottomrule
  \end{tabular}}
  \label{tab:lightweight_comparison}
\end{table}

\begin{figure}[t]
  \centering
  \subfigure[Manga109 \cite{Matsui2017Sketch} with SR$\times$4]{\label{subfig:parmas_vs_psnr}
  \begin{minipage}[t]{0.48\textwidth}
    \centering
    \includegraphics[scale = 0.36]{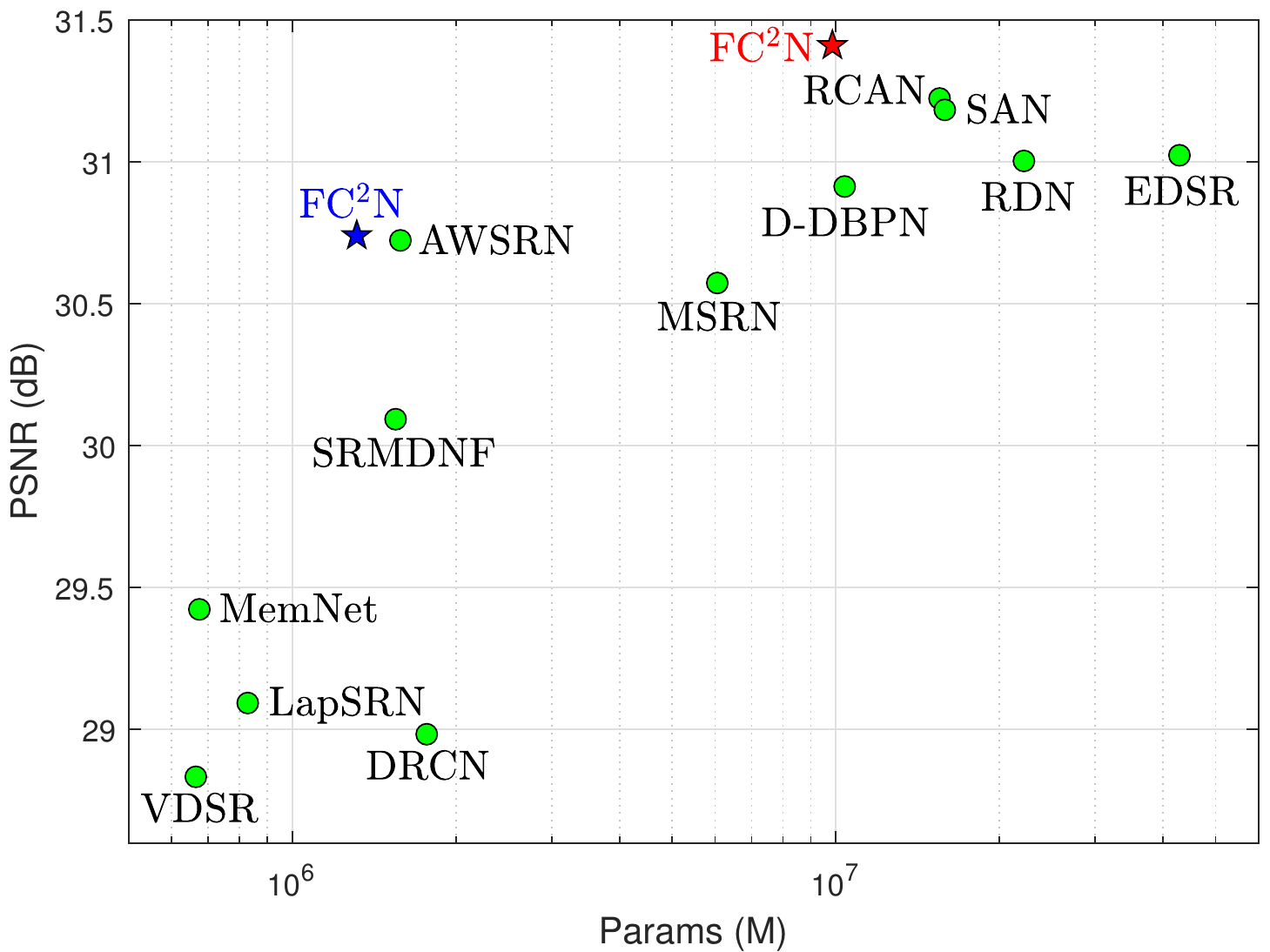}
  \end{minipage}}
  \subfigure[Set5 \cite{Bevilacqua2012Low} with SR$\times$2]{\label{subfig:multiadds_vs_psnr}
  \begin{minipage}[t]{0.48\textwidth}
    \centering
    \includegraphics[scale = 0.36]{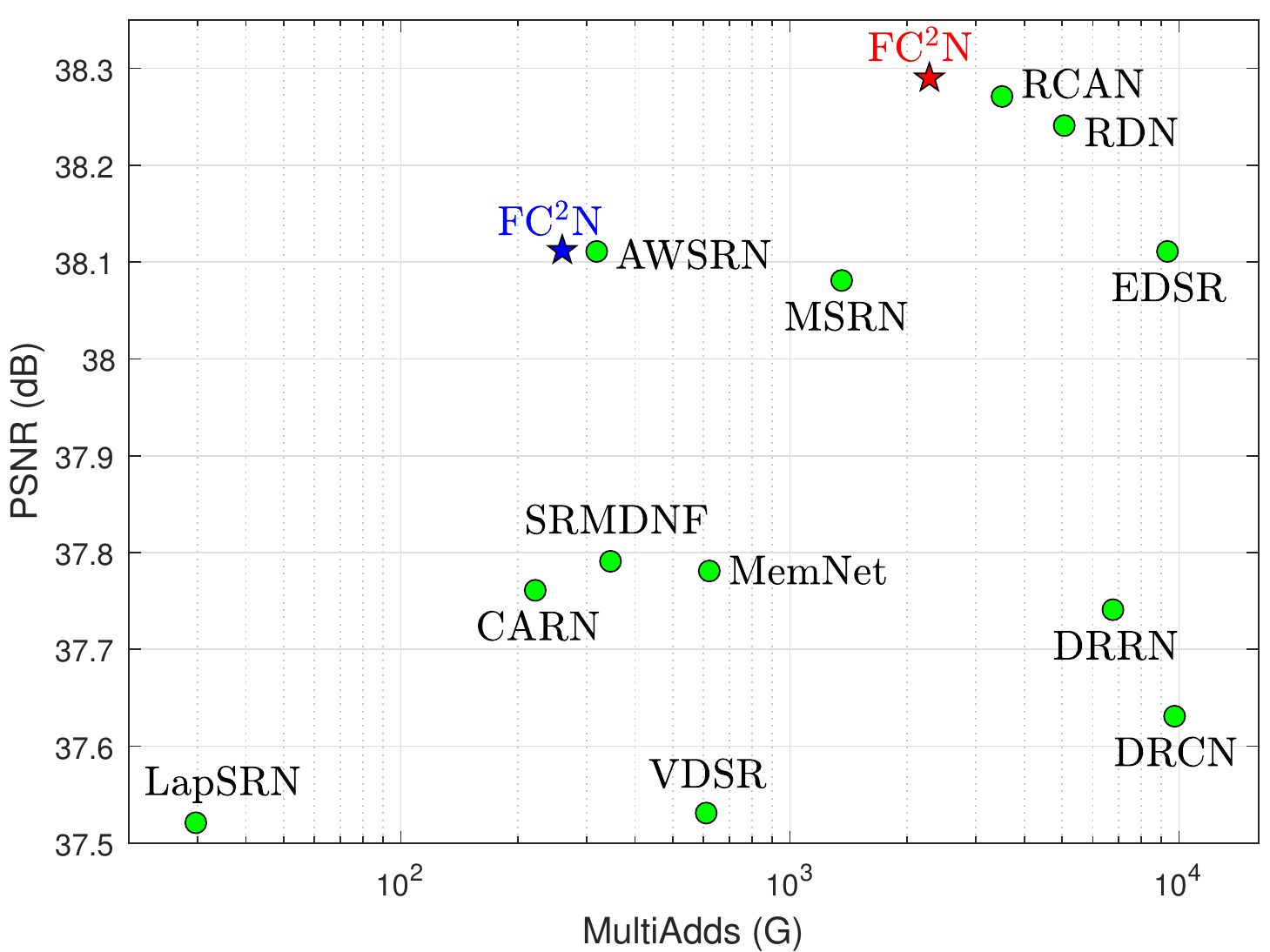}
  \end{minipage}}
  \vspace{-3mm}
  \caption{Efficiency analysis on parameters and computational overhead. {\textcolor[rgb]{1,0,0}{Red}} represents largescale implementation, {\textcolor[rgb]{0,0,1}{blue} denotes lightweight version.}}
  \label{fig:performance_analysis}
\end{figure}

\vspace{-3mm}
\subsection{Comparison with Advanced Methods}
\vspace{-2mm}
In this section, we compare both largescale and lightweight FC$^2$N models with other advanced methods. Similar to \cite{Lim2017Enhanced}, \cite{Zhang2018Residual} and \cite{Zhang2018Image}, we also use the geometric self-ensemble to improve performance, which is denoted as FC$^2$N$^{+}$. Besides, we introduce another strategy to further boost model performance, i.e., data range ensemble, which is computed as following: for each testing image $\mathbf{x}$, generate an image $\bar{\mathbf{x}}$ with complementary data range by $\bar{\mathbf{x}} = 255 - \mathbf{x}$, and then feed it into the model to produce HR output $\bar{\mathbf{y}}$. The final HR prediction of $\mathbf{x}$ is given by $[\mathbf{y} + (255 - \bar{\mathbf{y}})]/2$. When both strategies are applied, it is denoted as FC$^2$N$^{++}$.

\vspace{-4mm}
\subsubsection{Largescale Implementation}
Table \ref{tab:quantcomp} illustrates the quantitative comparison between the proposed FC$^2$N and other largescale SR models in case of largescale implementation ($n = 16, m = 8$). It can be observed that our FC$^2$N outperforms most of other methods on all datasets. In particular, FC$^2$N$^{++}$ further improves the performance of FC$^2$N$^{+}$ on the whole, which verifies the effectiveness of data range ensemble. When the scaling factor is 8, our FC$^2$N performs slightly worse than RCAN \cite{Zhang2018Image} but still better than other SR models. However, our FC$^2$N only uses about 60\% model parameters of RCAN \cite{Zhang2018Image}, indicating that it provides a better trade-off between model performance and network scale.

Fig.\ref{fig:vis_largescale_FC2N} shows the visual comparison between other SR methods and our FC$^2$N on two testing images from Urban100 \cite{Huang2015Single} with SR$\times$4. For ``img\_058'', most of previous methods generate blurring artifacts at the fringes, especially for those in the lower left parts of the cropped images. However, only the FC$^2$N generate the result much closer to the ground truth. For image ``img\_073'', the blurring effect in the results of other methods in texture region is more obvious, but our FC$^2$N can still produce the result that can imply the potential structure more clearly. These comparisons show the good representational capacity our FC$^2$N.

\vspace{-4mm}
\subsubsection{Lightweight Implementation}
To show that our FC$^2$N makes full exploitation of representational capacity, we also compare our lightweight FC$^2$N ($m = n = 4$) with other lightweight SR models, as shown in Table \ref{tab:lightweight_comparison}. The size of HR images is assumed to be 1280$\times$720 (i.e., 720p) for the calculation of MultiAdds, as in \cite{Ahn2018Fast} and \cite{Wang2019Lightweight}. We can see that our FC$^2$N gives the best trade-off between performance and model scale, and computational burden. It provides comparable performance to AWSRN \cite{Wang2019Lightweight} with fewer parameters and computational overhead.

\vspace{-4mm}
\subsubsection{Efficiency Analysis}
To further illustrate the efficiency superiority of our FC$^2$N to other models, we plot model performance versus parameters and MultiAdds in Fig.\ref{subfig:parmas_vs_psnr} and Fig.\ref{subfig:multiadds_vs_psnr} respectively. As can be seen, our FC$^2$N shows the best trade-off on parameter utilization and calculation efficiency for both largescale and lightweight implementations.

\begin{figure}[t]
  \centering
  \includegraphics[width=0.98\textwidth]{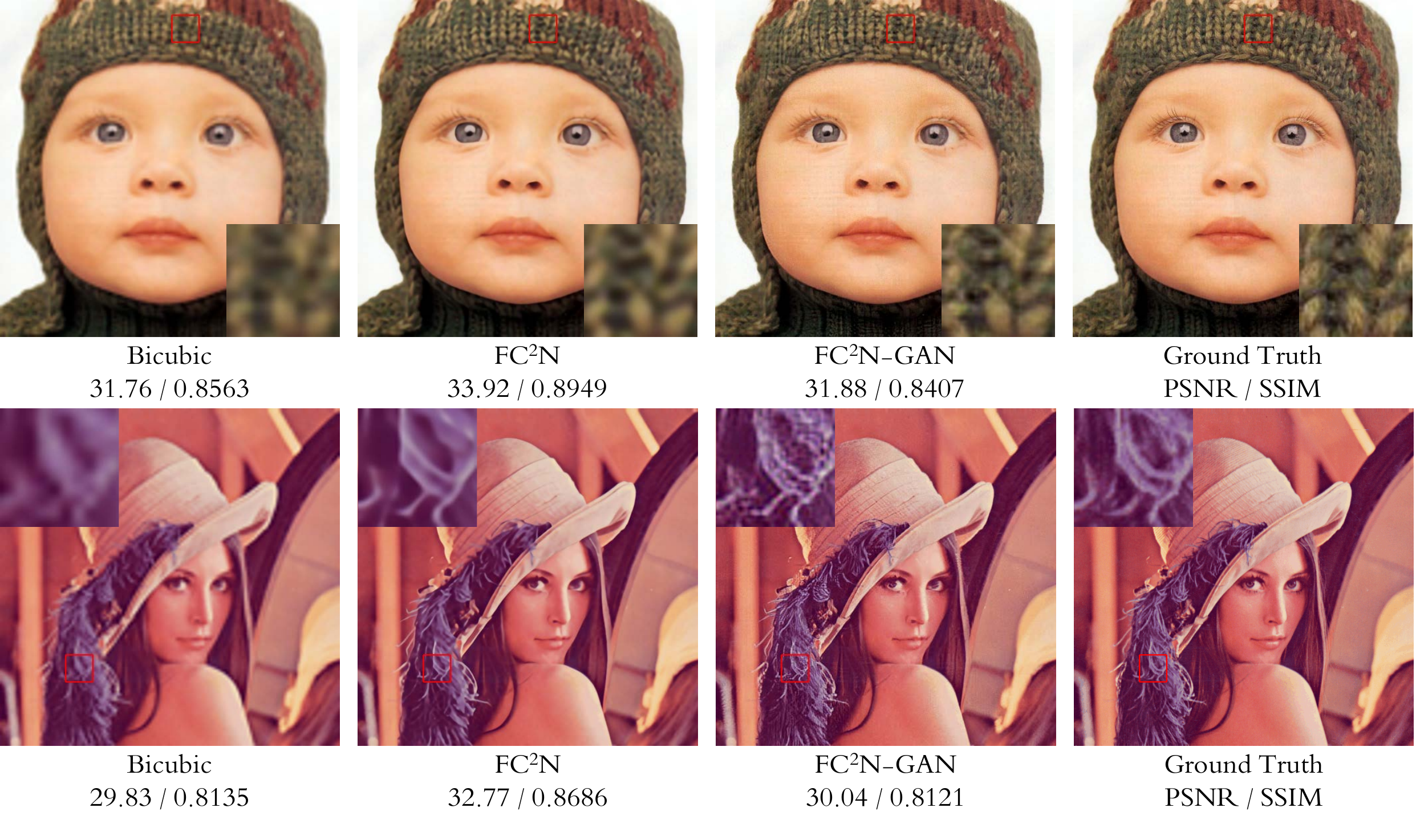}
  \vspace{-3mm}
  \caption{Visual comparison of largescale FC$^2$N with adversarial training ($n = 16, m = 8$)}
\label{fig:FC2N_GAN}
\end{figure}

\subsection{Adversarial Training}
The proposed FC$^2$N can also be extended to adversarial training \cite{Goodfellow2014Generative} for visually pleasuring SR results. We build a GAN model with our largescale FC$^2$N as the generator ($G$) and a 5-layer FCNN \cite{Shelhamer2017Fully} as the discriminator ($D$). The first three conv layers in $D$ have 3$\times$3 kernels, and the output channels are set to 64, 128 and 256 respectively. Then, a global average pooling is used to squeeze the spatial dimensions, which is followed by two 1$\times$1 layers that have 1024 and 1 output channels respectively. To train this GAN model, we adopt the same loss function for $G$ as \cite{Haris2019Deep,Haris2018Deep}, which has the form of $\lambda_1L_{mse} + \lambda_2L_{vgg} + \lambda_3L_{adv} + \lambda_4L_{sty}$. We set $\lambda_1 = 1.0, \lambda_2 = 0.1, \lambda_3 = 2 \times 10^{-3}, \lambda_4 = 1.0$ in our experiment.

Fig.\ref{fig:FC2N_GAN} shows a visual result for SR$\times$4. The model with adversarial training generates more plausible result on texture regions, even better than the ground truth (e.g., hair in the Lena image). However, the recovered results may not be faithfully produced with unexpected artifacts.

\vspace{-2mm}
\section{Conclusions}
\vspace{-2mm}
\label{sec:conclusion}
In this paper, we present a novel and simple network structure aimed at effective image SR tasks, i.e., FC$^2$N. Compared with previous advanced models, a major technical novelty of our FC$^2$N is the introduction of WCC as all skip connections in the network, and the avoidance to use residual learning. Through WCC, the model can not only adaptively select effective interlayer skips and make full use of hierarchical features, but also pay joint attention to the linear and nonlinear features. The CIC structure with CGs and CBs can also ease local representation learning and allow the model to fuse features from a fine to coarse level. Extensive experiments show that our FC$^2$N model outperforms most advanced models in both lightweight and largescale implementations, verifying its effective mining of model representational capacity.

%
\bibliographystyle{splncs04}
\bibliography{egbib}
\end{document}